\documentclass{article}

\usepackage{times}
\usepackage{geometry}
\usepackage{comment}
\usepackage{fancyhdr}
\usepackage{amsmath,amsthm,amssymb}
\usepackage{graphicx}
\usepackage{caption,subcaption}
\usepackage{color}
\usepackage{soul}
\usepackage[numbers,sort&compress]{natbib}

\title{Is Gun Violence Contagious?}
\author
{Charles Loeffler (University of Pennsylvania), Seth Flaxman (University of Oxford)}

\begin{document}

\maketitle

\begin{abstract}
{Existing theories of gun violence predict stable spatial concentrations and contagious diffusion of gun violence into surrounding areas. Recent empirical studies have reported confirmatory evidence of such spatiotemporal diffusion of gun violence. However, existing tests cannot readily distinguish spatiotemporal clustering from spatiotemporal diffusion. This leaves as an open question whether gun violence actually is contagious or merely clusters in space and time. Compounding this problem, gun violence is subject to considerable measurement error with many nonfatal shootings going unreported to police. Using point process data from an acoustical gunshot locator system and a combination of Bayesian spatiotemporal point process modeling and space/time interaction tests, this paper demonstrates that contemporary urban gun violence does diffuse, but only slightly, suggesting that a disease model for infectious spread of gun violence is a poor fit for the geographically stable and temporally stochastic process observed.}
\end{abstract}

\section*{Introduction}
Beginning with \cite{loftin_assaultive_1986}, scholars of crime have described homicide in general, and gun-related homicide in particular as diffusion processes with contagion or epidemic properties \cite{reiss_understanding_1993,fagan_social_2007,patel_contagion_2012}. This criminological and public health perspective holds that violence  begets more violence and provides a theoretical foundation for policy interventions such as violence interruption programs \cite{skogan_evaluation_2009,webster_effects_2012}, designed to disrupt the transmission of gun violence. 
\par
Empirical studies of the spatiotemporal properties of crime have reported robust evidence of space-time clustering of property crime \cite{johnson_repeat_2008,mohler_self-exciting_2011, short_dissipation_2010,townsley_infectious_2003} and violent crime \cite{cohen_diffusion_1999,morenoff_neighborhood_2001,tita_measuring_2004,ratcliffe_near-repeat_2008,braga_concentration_2010,rosenfeld_facilitating_1999}. Cohen and Tita \cite{cohen_diffusion_1999,tita_measuring_2004} report evidence of spatial spillover of homicide and shots fired from census tracts with increasing rates to adjacent tracts, but only during peak crime periods. Messner et al. \cite{messner_spatial_1999} report non-random clustering of homicides at the county level, and Ratcliffe and Rengert \cite{ratcliffe_near-repeat_2008} reports evidence of a non-random spatiotemporal clustering at the block-level. Both interpret non-random clustering as potential evidence of violence diffusion. However, scholars have yet to determine whether gun violence actually diffuses in space and time, consistent with an epidemic or similar contagion process, or merely clusters in space and time, consistent with endemic gun violence concentrated in certain locations at certain times. This limitation is due, at least in part, to the fact that existing space/time tests were designed to detect departures from complete spatiotemporal randomness rather than distinguish between different types of non-random clustering. 
\par
Given the high levels of observed gun violence in many U.S. cities \cite{braga_concentration_2010}, including retaliatory shootings embedded in social networks that are themselves embedded in neighborhoods characterized by concentrated poverty \cite{morenoff_neighborhood_2001,tita_impact_2007,papachristos_murder_2009}, gun violence could conceivably become contagious with violence triggering more violence \cite{fagan_social_2007,patel_contagion_2012}. Lending empirical support for this possibility, other human behaviors, including self-directed violence, have been shown to diffuse readily \cite{tarde_laws_1903,coleman_diffusion_1957,christakis_spread_2007} and evidence of diffusion of violence in social networks has been recently reported \cite{rosenfeld_facilitating_1999, papachristos_murder_2009,short_gang_2014,papachristos_2016}. 
\par
At the same time, detailed examinations of homicide circumstances and motivations have revealed that much fatal gun violence is spontaneous \cite{metropolitan_police_department_report_2006,philadelphia_police_department_murder/shooting_2014,chicago_police_department_2011_2012}, resulting from arguments and failed drug transactions (See also \cite{rand_reducing_2003}. The existence of non-retaliatory shootings alone, if they were randomly distributed in space-time, would simply add noise to estimates of the diffusion of any retaliatory shootings. However, if a sufficient fraction of all gun discharges in a city result from non-retaliatory shootings that are themselves clustered in space-time, then endemic gun violence with little or no diffusion in space or time could be confused for an epidemic or infectious process using conventional space/time interaction tests. 

\par
To test these alternative possibilities and to more precisely describe the spatiotemporal point process of gun violence, we examine a unique dataset composed of data from an acoustic gunshot locator system (AGLS) installed in Washington, D.C. Using this data, which overcomes the incomplete measurement of gun violence in conventional reported crime data, and refined space-time interaction tests, we observe that while gun violence does diffuse over space and time, this diffusion is quite minimal---limited in space to 126 meters and in time to 10 minutes--and thus much more likely to be consistent with a discrete gun fight, lasting for a matter of minutes, than with a diffusing, ``infectious'' process linking violent events across hours, days, or weeks. As such, these results provide little support for the classic contagion models of violence diffusion; instead, this finding supports models predicting stochastic clustering of gun violence in space and time \cite{braga_concentration_2010}.

\section*{Methods}
At its most basic level, gun violence is an example of a spatiotemporal point process. Individual gunshots occur at specific geographic coordinates and at specific times. As such, gun violence lends itself to analysis using one of the several space-time interaction tests that have been developed to assess whether a point process is consistent with the random allocation of points across time and space or is sufficiently  clustered that further investigation is warranted. One of the earliest space-time interaction tests, proposed by Knox \cite{knox_detection_1964,knox_epidemiology_1964}, examines the distribution of spatiotemporal distances between all points, reducing this combinatorial set of pairwise relationships to a simple 2x2 contingency table of near and far distances in both dimensions. Too many points in the near in time and space cell (compared to a null model of independence) indicates clustering, which is not consistent with a homogeneous Poisson process. Ratcliffe and Rengert \cite{ratcliffe_near-repeat_2008} recently analyzed Philadelphia shooting data using this test and observed that an elevated number of shootings was present within two weeks and one block of a prior shooting. 
\par
Tests for space-time interaction like the Knox test are really testing for spatiotemporal clustering \cite{mantel_detection_1967}, yet spatiotemporal clustering could have various explanations. Gunfire that occurs mostly on weekend nights in high crime neighborhoods, for instance, would produce reliable spatiotemporal clustering despite the fact that the spatial and temporal processes generating the shootings are independent. The difficulty of distinguishing between a non-diffusing but heterogeneous point process such as this and a diffusing point process in which each new point is potentially the parent of additional points in its immediate vicinity was identified as early as \cite{bartlett_spectral_1964}. More recently, \cite{diggle_spatial_2013}~reiterated the importance and the difficulty of separating these two types of processes in a range of applications where stable and segregated point processes suggest substantially different theoretical mechanisms. 
\par
Up until recently, scholars in criminology and spatial statistics have used non-parametric descriptive statistics to assess the departures from randomness observed in the spatiotemporal crime processes. For example, \cite{wooditch_2015}, demonstrated the uses of Ripley's K-Function for simultaneously examining space and time in program evaluations of stop and frisk in New York City. However, non-parametric spatial statistics have only recently been adapted to handle inhomogeneous underlying intensities, which are the norm in criminological applications. And it remains an open question whether these methods will prove robust enough to handle the typical criminological case in which the underlying intensity of crime is unknown. For this reason, scholars in a number of fields have adapted model-based spatiotemporal methods for use in characterizing spatiotemporal processes with both underlying intensities and conditional intensities. In work on crime, Mohler and colleagues were the first to demonstrate the utility of these models to the problem of predicting spatiotemporal point processes \cite{mohler_self-exciting_2011}. They showed that the Hawkes process model with its two components could readily fit the heterogeneous spatiotemporal point process of crime in urban settings while allowing for near-repeats events in the vicinity of high intensity property crimes. More recently, Meyer et al.~ \cite{meyer_2016}, showed that these same methods could serve to characterize a range of background and conditional intensities, in their application, hospital admissions. Building on these earlier efforts, we sought to adapt these methods for the problem of separating inhomogeneous background intensities from conditional intensities.

\paragraph*{Our model}
Modern spatiotemporal point process models and increasingly fine-grained spatiotemporal data allow us to distinguish between spatiotemporal clustering and diffusion.
To do this, we employ a Hawkes process model. Given a set of observations indexed by their space-time locations, $(x_1,y_1,t_1), \ldots, (x_n,y_n,t_n)$, the conditional intensity of the Hawkes process is given by:
\begin{equation}
	\lambda(x,y,t) = m_0 \cdot \mu(x,y,t) + \theta \sum_{i: t_i < t} \omega \exp\left(-\omega(t-t_i)\right) \frac{1}{2\pi\sigma^2}\exp\left(-((x-x_i)^2 + (y-y_i)^2)/(2\sigma^2)\right)
\label{eq:hawkes}
\end{equation}
where we have specified an exponential form for the temporal triggering kernel
and a Gaussian form for the spatial kernel following \cite{mohler_marked_2014}.
There are two components to this model, an
underlying {\em endemic} spatiotemporally varying intensity $\mu(x,y,t)$ weighted by a
non-negative parameter $m_0$ which we estimate and a
{\em self-excitatory} conditional intensity that depends on the history of
previous events.  We perform Bayesian inference to infer posteriors over the
parameters in the model. 
Priors and model implementation details are discussed in the Appendix in Section \ref{section:hawkes}.

Of particular interest are the posterior
distributions over the lengthscale (bandwidth) parameters for the spatial and temporal
kernels as these enable us to characterize the extent of diffusion in space and time. 
Also of interest is the excitation parameter $\theta$, which gives the average
number of shootings triggered by any particular shooting.
Finally, for a shooting at location $(x_i,y_i,t_i)$ we can calculate the ratio
$r_i = m_0 \cdot \mu(x_i,y_i,t_i) / \lambda(x_i,y_i,t_i)$. 
This ratio is the fraction of intensity explained
by our model's endogeneous component for observation $i$, while $1-r_i$ 
is the fraction explained by the excitatory component. Some points will be
better explained by the background process than others, and we explore
how much variation there is in $r_i$ for $i = 1, \ldots, n$.

To capture the first-order space/time clustering corresponding to the highly
uneven distribution of gun violence in space \cite{braga_concentration_2010} and
time \cite{metropolitan_police_department_report_2006,philadelphia_police_department_murder/shooting_2014,chicago_police_department_2011_2012},
we use kernel intensity estimation to separately estimate $\hat\mu_s(x,y)$ and $\hat\mu_t(t)$,
obtaining a separable estimator $\hat\mu(x,y,t) = \hat\mu_s(x,y)\hat\mu_t(t)$. This
separable approach to estimating the underlying intensity follows \cite{diggle2005point} where it
was used for an infectious disease modeled by a Cox process.
We are protected against overfitting
because $\hat\mu_s(x,y)$ and $\hat\mu_t(t)$ are estimated separately. We consider this to be a conservative approach
to take, insofar as overfitting might lead us to miss a weak self-exciting signal if we were to incorrectly
attribute it to the endemic background process.
We note that while it might be reasonable to include covariates like socioeconomic indicators from the census in this estimation step, no covariates match the spatiotemporal resolution of our data, so we would not expect them to add any additional information.
We use Epanechnikov kernels \cite{ramlau1983smoothing} with relatively long spatial and temporal
bandwidths. We perform a sensitivity analysis in the Appendix in Section
\ref{section:smoothing-kernel} showing that our results are robust to various choices 
of the bandwidth parameters of the smoothing kernel.
\par
While this two-stage model necessarily involves more complexity than past space/time clustering methods such as the Knox or the K-function, the segmentation of clustering into first-order and second-order components offers a convenient metric for evaluating the relative importance of background and conditional  intensities. This approach can also be thought of as a compromise between the non-model-based test statistics, which are ill-equipped to accommodate inhomogeneous backgrounds, and SIR models, which cannot easily be made to fit our setting, as there is no natural way of observing who within the population is susceptible to infection, infected, or recovered.

\section*{Data}
\begin{figure}[ht!]
	\centering
	\begin{subfigure}[t]{.5\linewidth}
                \includegraphics[width=\linewidth]{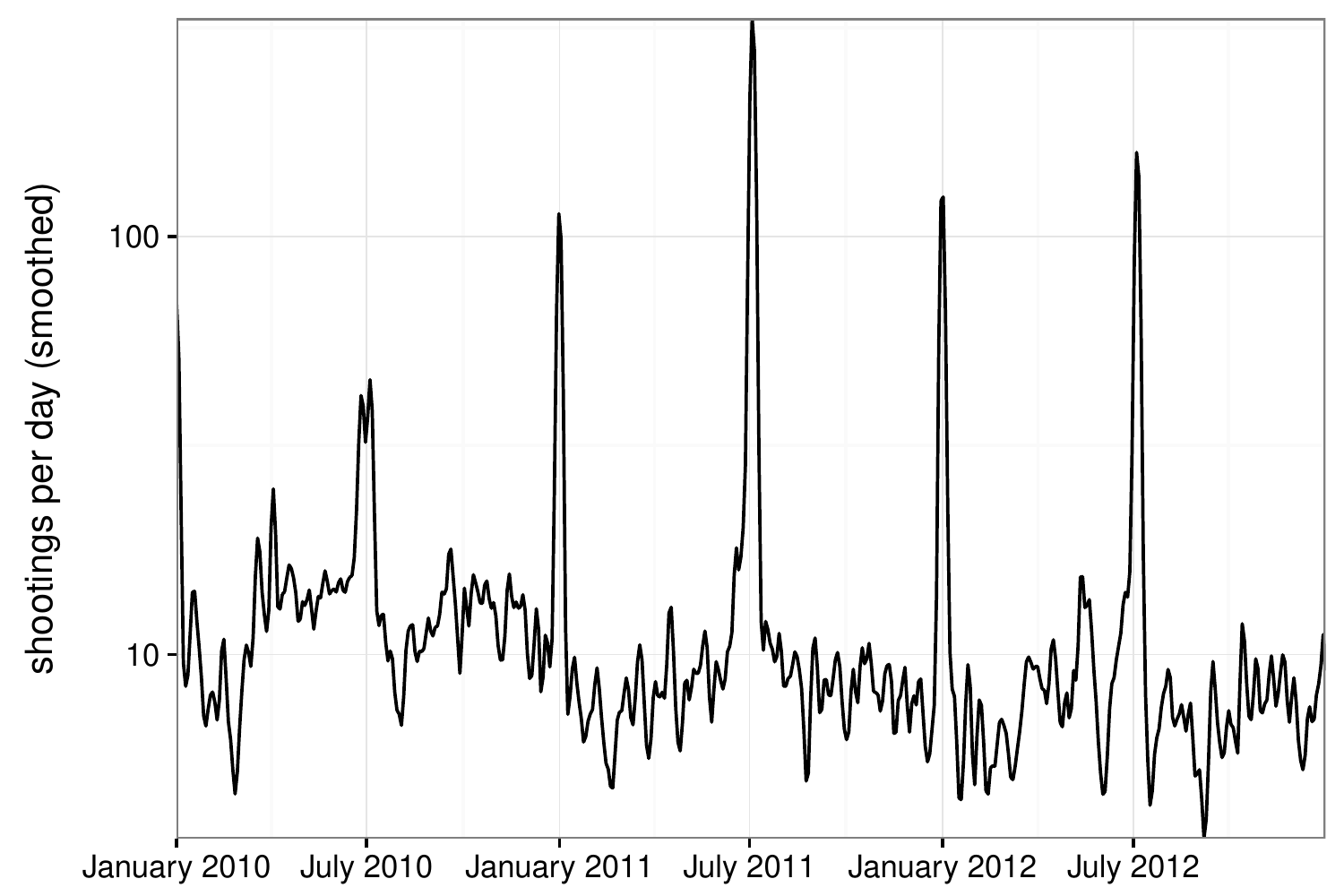}
                \caption{Time series}\label{fig:temporal-intensity-agls}		
            \end{subfigure}%
	\begin{subfigure}[t]{.5\linewidth}
		\centering
		\includegraphics[width=\linewidth]{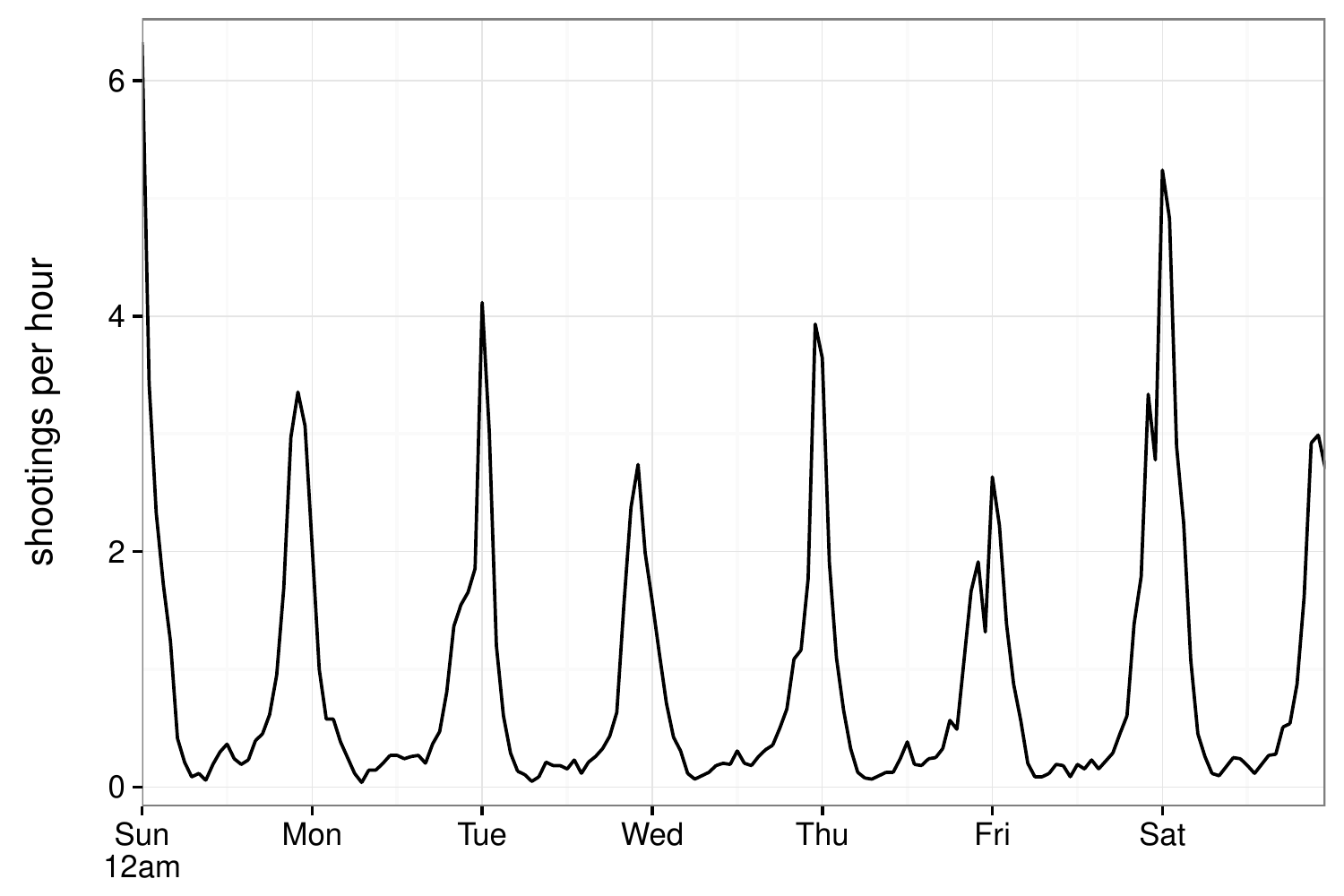}
                \caption{Hour of week trends}\label{fig:hourly-intensity-agls}
	\end{subfigure}
	\begin{subfigure}[t]{.5\linewidth}
		\includegraphics[width=1\linewidth]{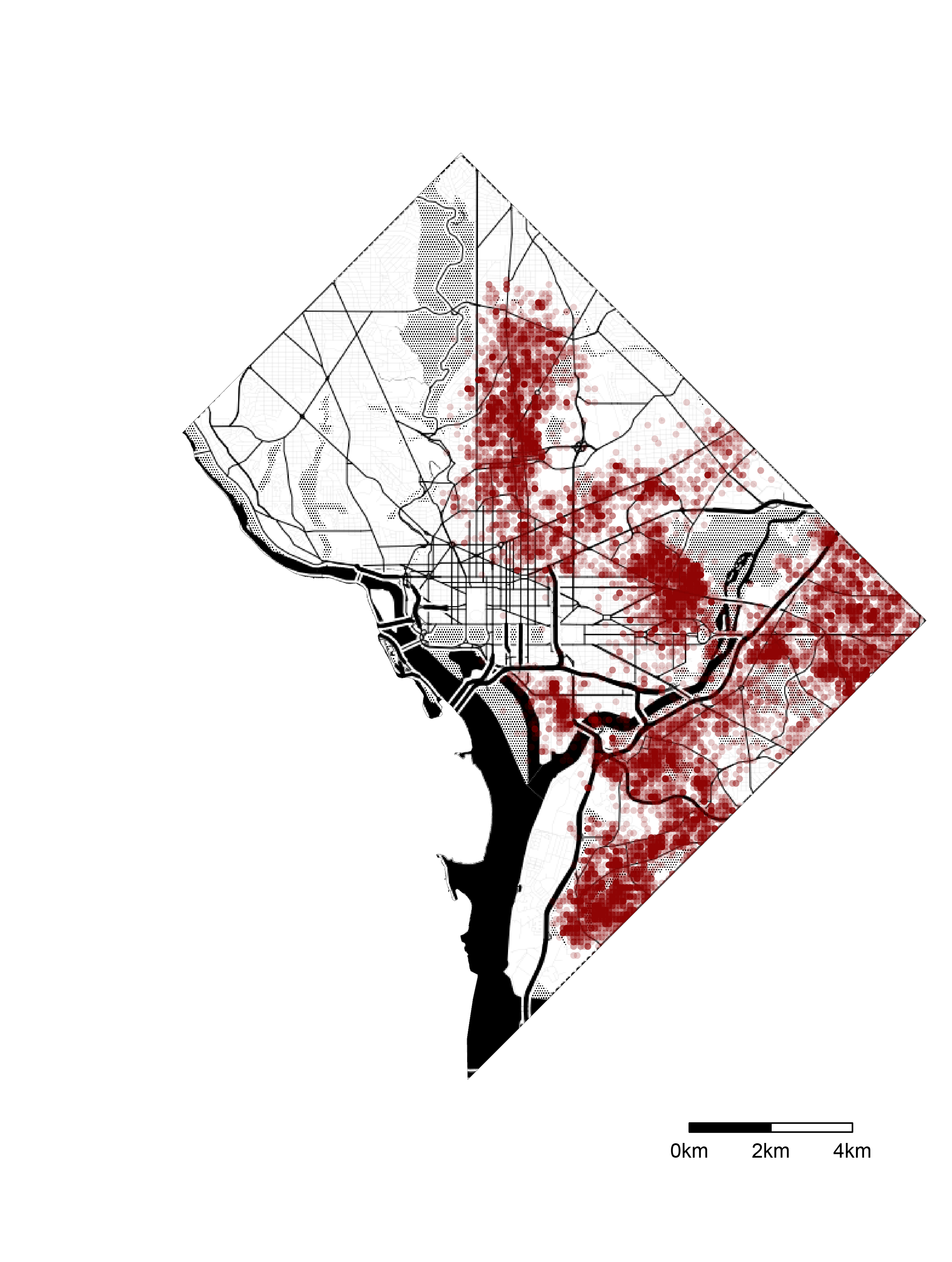}
            \caption{AGLS}
	\end{subfigure}%
	\begin{subfigure}[t]{.5\linewidth}
    		\includegraphics[width=1\linewidth]{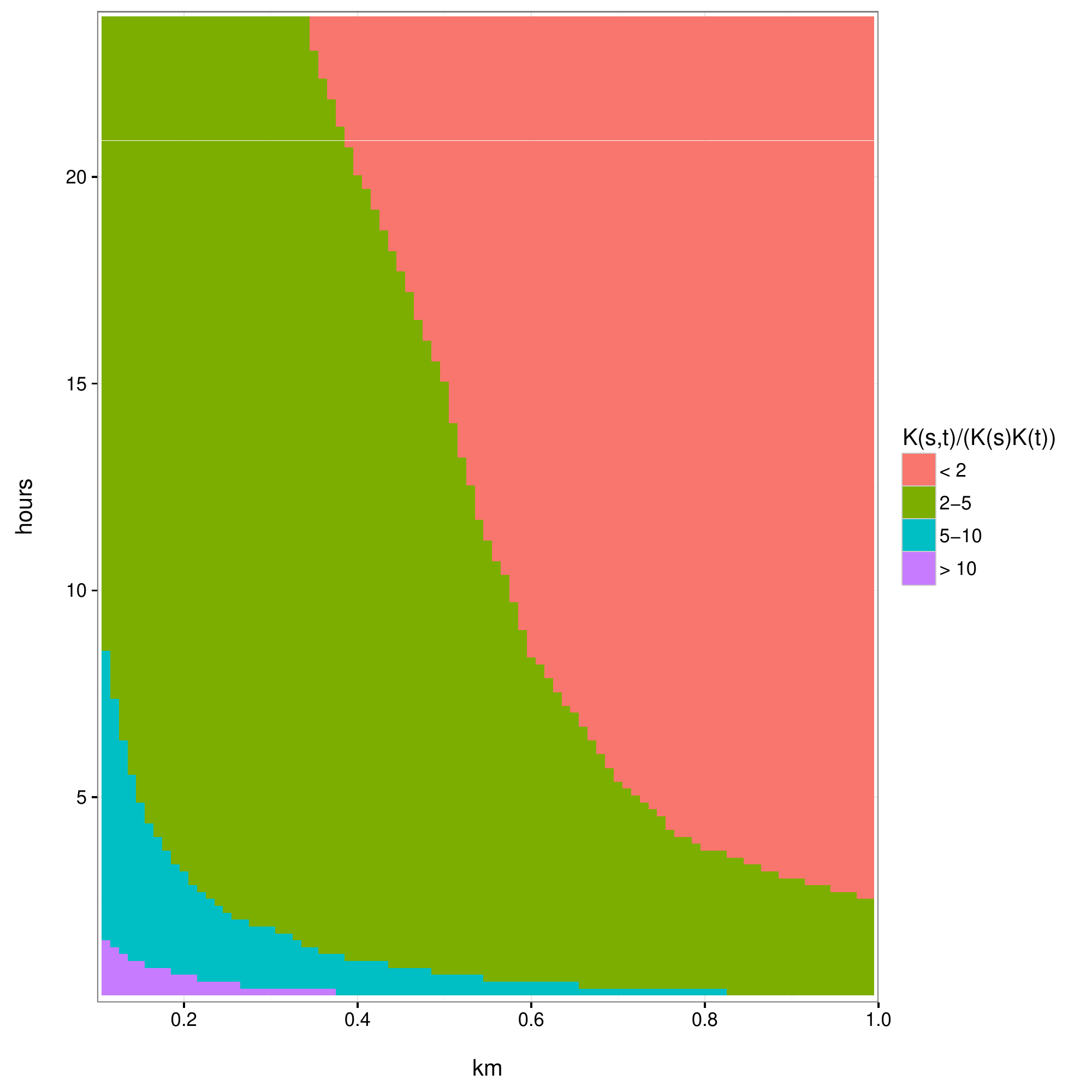}
                \caption{Ripley's K-Function}
            \end{subfigure}
        \caption{(a-b). Temporal distribution of acoustically located gunshots in Washington D.C., 2010-2012. (c). Spatial distribution of AGLS, 2010-12. (d). Ripley's K-Function.}
	\label{fig:spatiotemporal}
\end{figure}

Conventional crime data on gun violence is gathered from human-witnesses who notify police of fatal, non-fatal, and miss or near-miss shootings in their immediate vicinity. While reported shootings are generally thought to be a near-random subset of all discharges \cite{zimring_is_1967}, it is generally understood that no more than a quarter to a third of illegal firearm discharges are reported to police \cite{mares_evaluating_2012}. Due to concerns about this ``dark figure'' of unreported gun crime, increasingly, municipalities with persistent gun violence problems are deploying acoustical gunshot locator systems (AGLS) to enable the detection of illegal gun discharges unreported to police through 911 systems.
\par
A network of acoustical sensors, AGLS systems operate by recording the impulsive sounds of gun discharges and filtering out other impulsive sounds (e.g., jack-hammers, car-back-fires, fireworks) that might be confused with gunshot events. While military applications of this technology can use a combination of shockwave and muzzle blast features to detect and localize the point of origin of a ballistic projectile, the smaller caliber munitions commonly observed in domestic security applications preclude the use of the shockwave feature \cite{sallai_weapon_2011}. As a result, AGLS used in domestic public safety applications rely solely on time-of-arrival of the muzzle blast acoustical signal at multiple microphones arrayed at known distances. Using this information, the bullet origin can be computed using multilateration. While accuracy concerns remain, in practice, AGLS systems record between 2 and 4 times more firearm ``shots fired'' incidents than conventional 911 incident systems \cite{mares_evaluating_2012}.
\par
We use AGLS data from Washington, D.C. covering the period from January 1, 2010 to December 31, 2012. Sensor coverage during this period included parts of six of the D.C.'s seven police districts with greatest sensor concentration in the Northeast and Southeast quadrants. Gunshot event data were rounded to produce approximately 100m spatial resolution and 1 second temporal resolution. For a detailed discussion of the DC deployment, see \cite{petho_acoustic_2013}. Shootings occurring less than 1 minute and 100m in time and space were merged (See Appendix Section \ref{section:duplicates}).  
\par
Figure \ref{fig:spatiotemporal}a indicates a high degree of temporal clustering around New Year's and July 4th. Removing these temporal clusters (see Appendix \ref{section:holidays}) does little to reduce seasonal temporal clustering, since summer months have much more activity than other months. High degrees of day-of-week and hour-by-hour periodicity can also be observed (Figure \ref{fig:spatiotemporal}b). The spatial intensity of acoustically detected gunshots (Figure \ref{fig:spatiotemporal}c) also demonstrates a high degree of spatial clustering. Noticeable concentrations can be observed in subsections of the southeast and northeast city quadrants. Adjacent areas appear to have sparser densities. 
 \par
 Compared to conventional firearm use data, as seen in Appendix Figure \ref{fig:intensity}c, the AGLS firearm data shows much higher densities of illegal firearm usage. For the city as a whole, during the years 2011 and 2012, police recorded 577 and 614 firearm assaults, respectively. By comparison, the AGLS system, despite incomplete coverage, recorded 6,668 and 5,385 gunshot events in 2011 and 2012, respectively. These striking level differences reflect both the considerable number of unreported discharges as well as the sizable number of miss events that did not result in physical injury. 
\par
This considerable amount of spatial and temporal clustering means that the null hypotheses for classical tests of space/time clustering will always be rejected. For the present application, following \cite{diggle_second-order_1995}, we calculated Ripley's K function in space and time, $K(s,t)$ and divided it by the product of separate K functions $K(s)K(t)$. $K(s,t)$ is proportional to the expected number of additional points within a distance $s$ and time window $t$ of an arbitrary event \cite[p. 210]{cressie2011}. By comparing $K(s,t)$ to $K(s)K(t)$ we are looking for excess gunshots as compared to how many we would expect based on the separable product of $K(s)$ and $K(t)$. Based on this analysis, space-time clustering is clearly evident (see Figure \ref{fig:spatiotemporal}d).  If firearm assault or homicide data approximate a random sample of all firearm discharges, this inhomogeneous clustering could be observed in conventional crime data as well.
\begin{figure}[ht!]
	\centering
	\begin{subfigure}[t]{.5\linewidth}
		\centering
            \includegraphics[width=\linewidth]{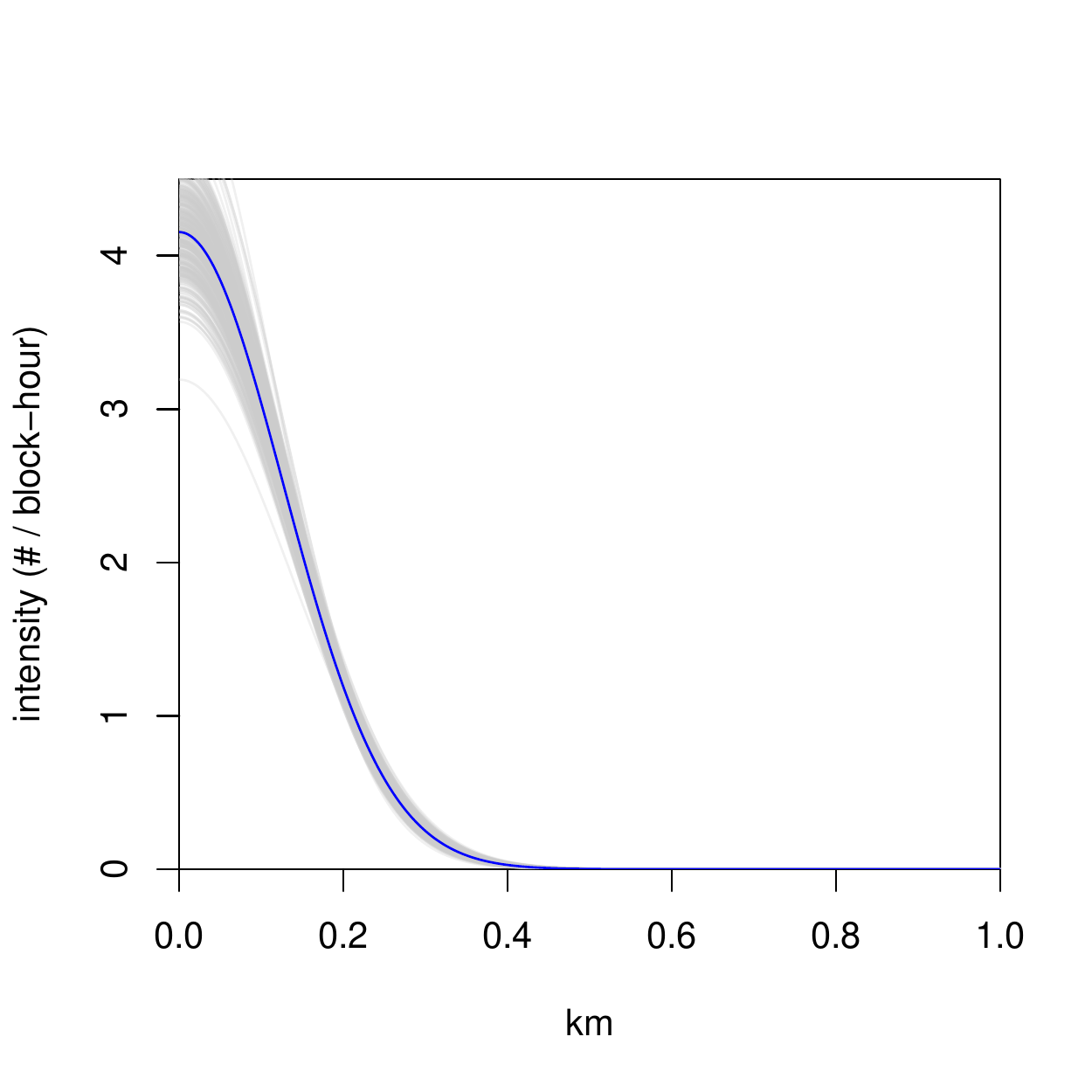}
            \caption{Conditional intensity in space at time $t = 0$}
	\end{subfigure}%
	\begin{subfigure}[t]{.5\linewidth}
		\centering
                \includegraphics[width=1\linewidth]{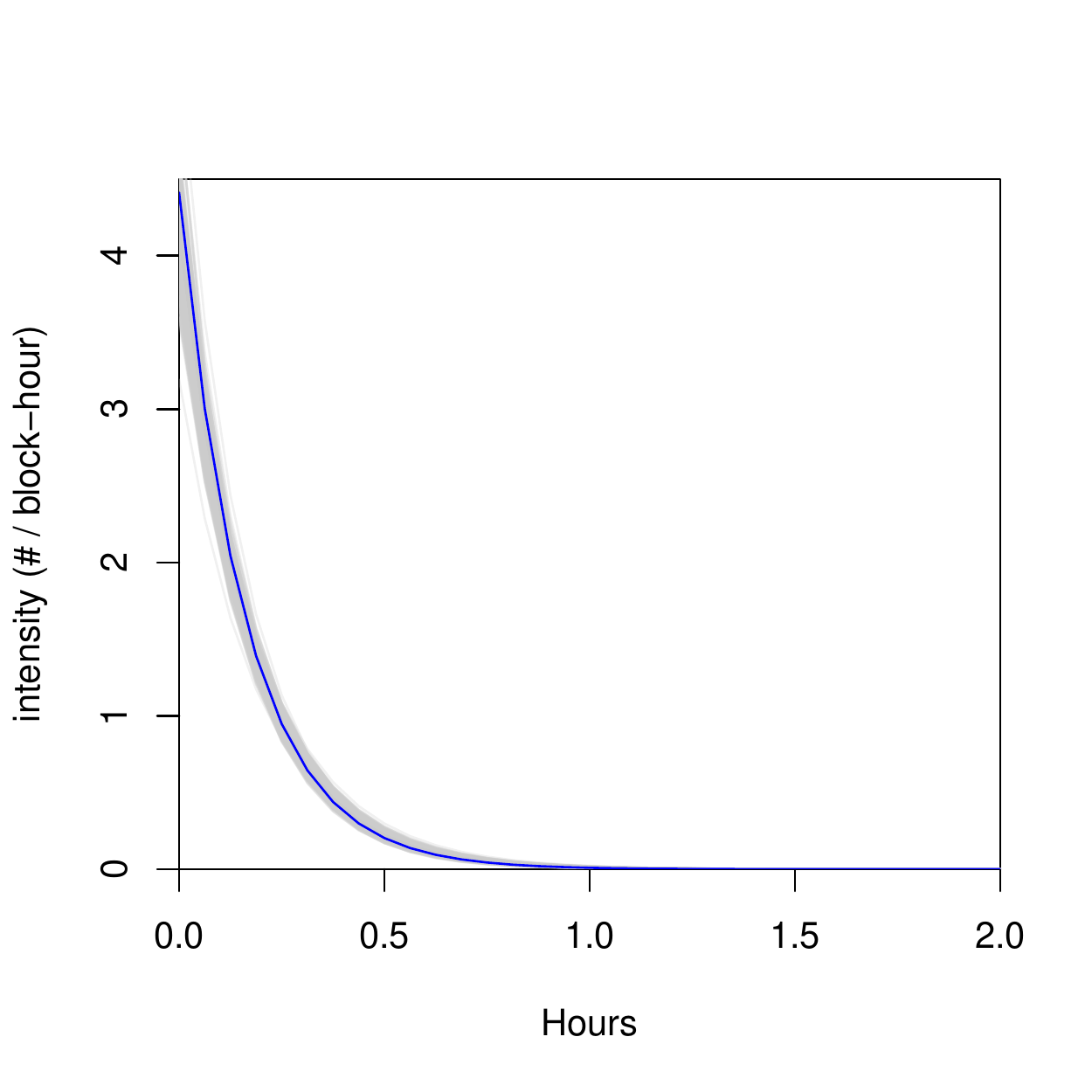}
            \caption{Conditional intensity in time at distance $s = 0$}
	\end{subfigure}%

	\begin{subfigure}[t]{.5\linewidth}
        \includegraphics[width=1\linewidth]{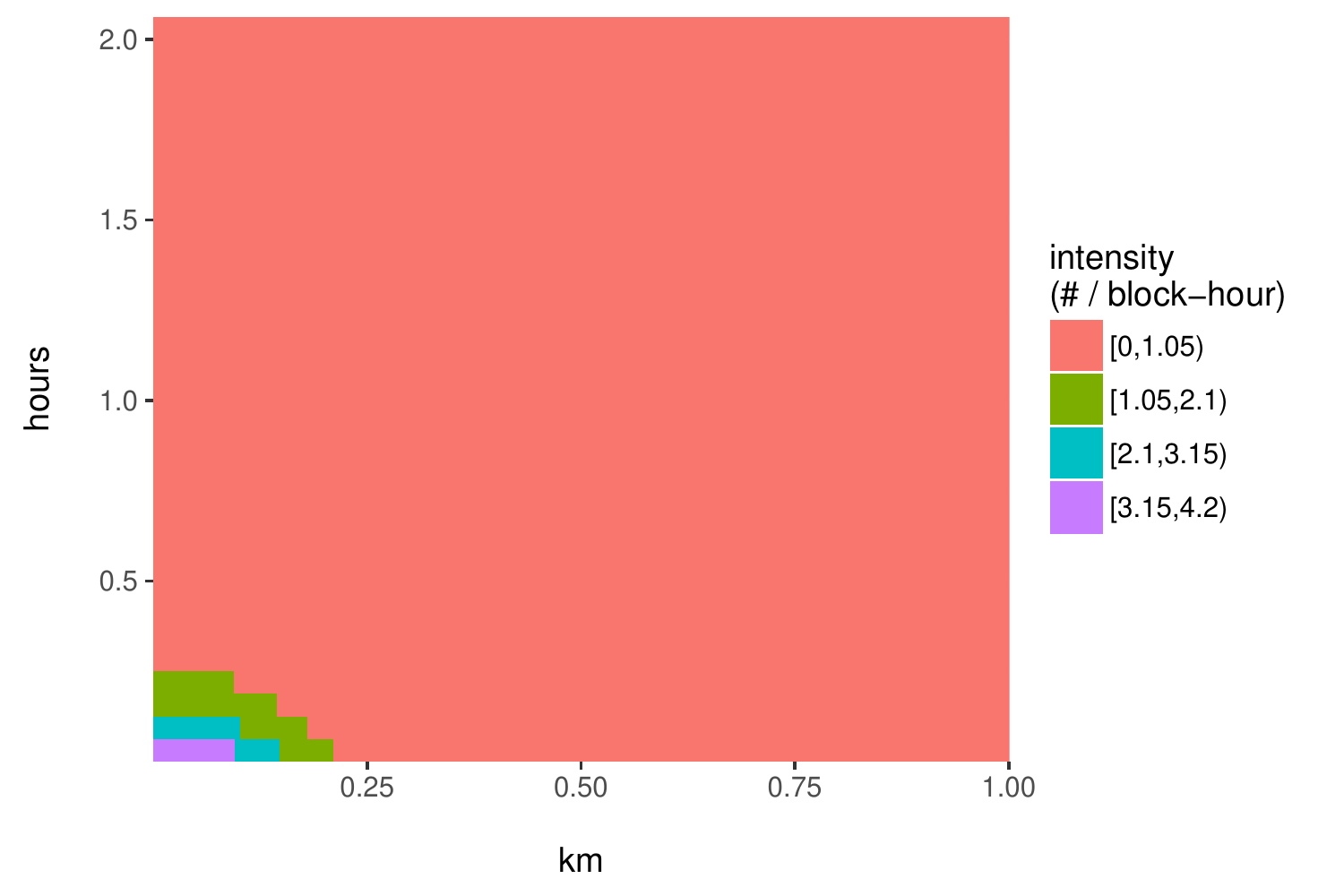}
            \caption{Spatiotemporal conditional intensity}
    \end{subfigure}
    \caption{The self-excitatory component of the Hawkes process, corresponding to the product of a spatial and temporal kernel,
        is visualized in space (a) at time $t=0$, in time (b) at distance $s=0$, and in space/time (c). In (a) and (b)
        the blue line is the mean of the posterior distribution. The gray lines show all 400 samples from the posterior distribution,
    reflecting the (small amount) of posterior uncertainty in the parameter estimates. }
    \label{fig:triggering-kernel}
\end{figure}

\section*{Results}
We estimated the underlying spatiotemporal intensity by separately performing kernel intensity estimation
in space and time, with a spatial bandwidth of 1.6 kilometers and a temporal bandwidth of 14 days. 
To perform Bayesian inference we used Markov Chain Monte Carlo (specifically HMC as implemented in 
Stan \cite{stan-software:2016}), running 4 chains for 200 iterations each, discarding 100 iterations as warmup, 
to obtain a total of 400 draws from the posterior. Convergence diagnostics are in the Appendix in Section \ref{section:mcmc}.

The posterior estimates of the parameters in our model in Eq.~\eqref{eq:hawkes}
are as follows: $m_0$, the weight placed by our model on the underlying endogeneous intensity (as opposed to the triggered self-excitatory
component) was 0.87 [95\% posterior credible interval 0.85, 0.89]. Correspondingly, the model estimated
the parameter $\theta$, giving the average number of shootings triggered by a particular shooting,
to be 0.13 [0.12, 0.13]. Intriguingly, this split (87/13) between endogeneous and self-excitatory shootings closely matches the reported split (86/14) between non-retaliatory and retaliatory shootings from examining homicide motives \cite{metropolitan_police_department_report_2006}.
\par The temporal lengthscale $1/\omega$ for the exponential triggering kernel is 
estimated to be 10 minutes [9.5,11]. The spatial lengthscale $\sigma$ was estimated to be 126 meters [121, 134].
The excitatory component of Eq.~\eqref{eq:hawkes} is visualized in Figure \ref{fig:triggering-kernel} where the extremely local
ranges of the self-excitatory process are evident. At time 0 and spatial distance 0, the excitatory component of
the Hawkes process has a value of 180 [158, 200] shootings per km$^2$-hour which equals 4.2 [3.7, 4.7] shootings per block-hour. Due to the extremely local range of the space and time kernels, the intensity drops off very quickly, so it is more intuitive to interpret to $\theta$. 

\begin{figure}[ht]
 \includegraphics[width=\linewidth]{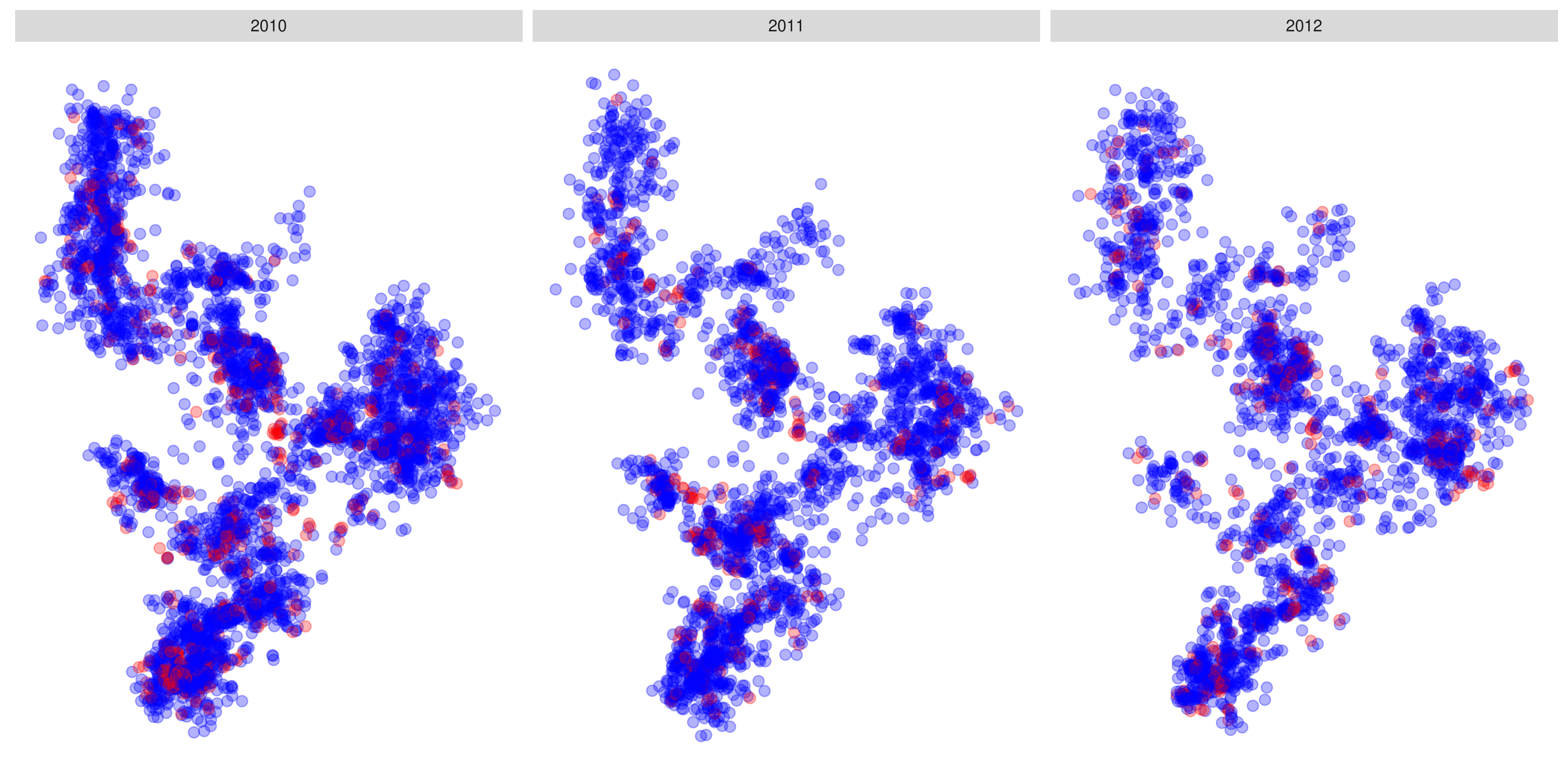}
 \caption{Background (blue) versus ``triggered'' (red) shootings, by year. At any location in space-time, our inferred model yields
	 a posterior distribution over the intensity at that location. For each shooting in our dataset, we calculated the excitatory intensity (see Eq.~\eqref{eq:hawkes}) as a way of classifying shootings as either coming from a background process or as being ``triggered'' by previous shootings. Based on the parameter $\theta$ in our model, 13\% of shootings were caused by previous shootings, so we classified as ``triggered'' (red) shootings that fell in the top 13th percentile of excitatory intensity.}\label{fig:triggered}
\end{figure}

We interpret $\theta = 0.13$ as follows: if we were to observe 100 shootings at a given location (which would be many more than
we expect), we would on average expect to see 
13 additional shootings, very nearby in space and time (most within 11 minutes and 134 meters, using the upper end of the 95\% uncertainty interval). From those 13
additional shootings, we would expect to see be between 1 and 2 additional shootings, again very nearby in space and time. And so on,
for a total of about 115 shootings. Allowing for the most liberal spatial and temporal extents, the 15 offspring shooting
triggered by the original 100 shootings will occur within just half an hour and 500 meters.
Since there were a total of 9,410 shootings in our dataset, we thus considered the 1,255 shootings with the highest
estimated conditional intensity to be triggered. These shootings are highlighted
in red, with background shootings in blue, in Figure \ref{fig:triggered}.

To evaluate the fit of our model, we use it to make predictions
of the number of shootings we expect to see in 1 km$^2$ by 1 hour balls centered on a Cartesian grid spanning our
obervations. On a given day this means we make $102 \times 24$ predictions. If we were to use just the kernel
smoothed endogeneous intensity, the correlation between the observed number of shootings and the kernel smoothed
intensity surface is 0.06, which simply reflects the long lengthscales we chose for this smoothing.
If we make predictions using the Hawkes process model that we fit, the correlation is 0.20.

As a check on our basic finding, we turn to the Knox test,
a much simpler test for space/time interaction, used in, e.g. \cite{ratcliffe_near-repeat_2008}. Our results
suggest that there this is very local and concentrated space/time clustering, so the Knox test should reject
the null hypothesis of independence at {\em any} practical spatial and temporal bandwidths, and indeed for
134 meters and 11 minutes, the Knox test resulted in a p-value $\leq 0.001$. But we also expect that since the
Knox test is cumulative, any other cutoffs will be significant as well, e.g. the Knox test with cutoffs 2 km
and 14 days has p-value $\leq 0.001$. This puzzle originally motivated our use of more refined models. Perhaps we could resolve it by simply considering both lower and upper bounds for the ``close in space'' and ``close in time'' cutoffs, following a suggestion of \cite{diggle2013statistical}, but this search over 4 parameters leads immediately to multiple testing problems. 

\section*{Discussion}
Existing theories of gun violence predict stable spatial concentrations and widespread diffusion of gun violence into surrounding communities, which should, in principle, make gun violence susceptible to a range of interventions. Past research, however, has yet to show that gun violence diffuses in either space or time due in part to the space-time interaction tests employed, which have not allowed for a characterization of the extent of any departure from complete spatiotemporal randomness in the space-time distribution of gun violence. The present investigation addressed this limitation by  employing space-time interaction models that support a separation of space-time interactions consistent with epidemic or contagion-like diffusion processes from more conventional and constrained clustered distributions. Based on these new methods, it appears that while gun violence does diffuse in both space and time, this diffusion is quite minimal with spatial diffusion limited to just 126 meters and temporal diffusion to ten minutes after an initial shooting incident. 
\par
Given the many studies of gun violence that have reported evidence of retaliatory shootings as well as other forms of gun violence diffusion, it is worth considering why there is so little evidence of gun violence contagion in the present study. One possible explanation is that previous analyses have been unable to separate highly clustered space-time violence into endemic and epidemic components due to the use of spatiotemporal statistics designed only to separate space-time clustering from complete spatiotemporal randomness. Another possible explanation is that in the more complete distribution of gun violence observed in this study, gun violence is both more spontaneous and less reactive than in the subset of shootings that result in a homicide or near homicide. A final possibility is that most observed gun violence is spontaneous rather than planned. 
\par
Lending support for this final possibility are recent examinations of perpetrator motives in a number of cities \cite{metropolitan_police_department_report_2006,philadelphia_police_department_murder/shooting_2014,chicago_police_department_2011_2012}, which show that most shootings  result from arguments and failed drug transactions, and are therefore more likely than retaliatory shootings to be self-extinguishing. This appears to be the case for Washington, D.C., where retaliatory gun violence appears to be a less frequent explanation for gun-related homicides \cite{metropolitan_police_department_report_2006}.   Whether or not other cities such as Los Angeles or Chicago, with larger and more established gangs, follow this pattern perfectly, it is worth recognizing that even in these cities a large fraction of gun violence is likely to be spontaneous and therefore not sufficiently stable to support interventions premised on contagion or diffusion models \cite{rosenfeld_facilitating_1999,skogan_evaluation_2009,butts2015cure}. Similarly, to the extent that gang-related violence in other cities fosters retaliatory violence that diffuses through social networks rather than euclidian space, future work must separate non-random network clustering from network diffusion \cite{shalizi_homophily}. Finally, macro-historical or cultural diffusion theories of violence, which predict an equilibrium change in the acquisition or utilization of firearms accompanying changes in illegal markets or social structure, will require somewhat different empirical tests that can adequately account for the multi-year time scale at which these mechanisms are theorized to operate \cite {blumstein_youth_1995,hemingway_2004}

\par
Taken together, these results suggest that the window of opportunity for reactive policy intervention is quite narrow. Interventions requiring more than a few minutes to implement, are likely to be too late to stop the diffusion of contagious violence. Likewise, interventions that target features of events beyond the elevated risk windows identified by this study are likely to have a difficult time discriminating between space-time locations that are still at elevated risk and those that have returned to no more than background risk levels. However, interventions that focus on locations with known stochastic clusters of gun violence, such as hotspot policing \cite{braga_hot_2005}, are more likely to be effective at suppressing it. Since, over very short time and spatial ranges, we find evidence that violence contagion explains a minority of shootings, and over longer time and spatial ranges we find no evidence for violence contagion, our results suggest that the metaphor of violence contagion may foster a misimpression \cite{dodge_framing_2008}, since it is does not accurately reflect a considerable fraction of typical urban gun violence. 

\newpage
\bibliographystyle{plainnat}
\bibliography{loeffler2}

\pagebreak
\appendix
\section{Appendix}
\subsection{Hawkes process}
\label{section:hawkes}
Given a temporal point pattern $t_1, \ldots, t_n \in \mathcal{R}$,
observed on a time window $[0,T]$,
the Hawkes process is a so-called self-exciting point process. It
has a parsimonious form, parameterized by a conditional intensity
function with two parts, a background rate $\mu_t(t)$ which may or may not vary with $t$
and a self-excitatory rate which is based only on events which occurred before time
$t$:
\begin{equation}
	\lambda(t) = \mu_t(t) + \theta \cdot \sum_{i: t_i < t} \omega \cdot \mbox{exp}(-\omega (t-t_i))
\label{eq:hawkes}
\end{equation}
Here the triggering kernel is given by an exponential with inverse lengthscale $\omega$
and weight $\theta$, which characterizes the strength and duration of an event's influence
on future events.
The likelihood of the Hawkes process is simply that of an inhomogeneous Poisson process
with intensity $\lambda(t)$:
\begin{equation}
\mathcal{L}(t_1, \ldots, t_n | \theta, \omega, \mu) = \prod_i \lambda(t_i) \exp(-\int_{t=0}^T \lambda(t) dt)
\end{equation}

To include a spatial dimension, we consider a spatiotemporal point pattern with
events at locations $(x_i,y_i,t_i)$ observed on a window $S \times [0,T]$. We further pick a spatial kernel $k(s,s')$,
and then we have the following conditional intensity:
\begin{equation}
\lambda(s,t) = \mu(x,y,t) + \theta \sum_{i: t_i < t} \omega \cdot \mbox{exp}(-\omega(t-t_i)) k((x,y),(x_i,y_i))
\label{eq:hawkes}
\end{equation}
The likelihood is as follows:
\begin{equation}
\mathcal{L}(s_1, t_1, \ldots, s_n, t_n | \theta, \omega, \mu) = \prod_i \lambda(s_i,t_i) \exp\left(-\int_S \int_{t=0}^T \lambda(s,t) ds dt\right)
\label{eq:hawkes-st-like}
\end{equation}
Then we can explicitly calculate
the log-likelihood corresponding to Eq.~\eqref{eq:hawkes-st-like}:
\begin{align}
	& \log \mathcal{L} = \sum_j \mu(x_j,y_j,t_j) + \theta \sum_{i: t_i < t_j} \omega \cdot \mbox{exp}(-\omega(t_j-t_i)) k((x_j,y_j),(x_i,y_i))  \\ 
	& - \int_S \int_{t=0}^T \left(\mu(x,y,t) + \theta \sum_{i: t_i < t} \omega \cdot \mbox{exp}(-\omega(t-t_i)) k((x,y),(x_i,y_i))\right) ds dt 
\end{align}
Now for simplicity, we assume that $\int_S k((x,y),(x_i,y_i)) ds = 1, ~~ \forall i$ which we
can ensure by properly normalizing the spatial kernel, and also assuming that the lengthscale is much shorter than the size of the spatial domain. 
We use a Gaussian kernel for $k((x,y),(x',y'))$ with lengthscale $\sigma$. Since we are in two dimensions, we have the following
normalized kernel:
\begin{equation}
k(s,s') = \frac{1}{2\pi\sigma^2} \exp(-\frac{\|s-s'\|^2}{2\sigma^2})
\end{equation}

We estimate the background intensity $\mu(x,y,t)$ using kernel smoothing with an Epanechnikov kernel:
\begin{equation}
	k(d) = \frac{3}{4}(1-d^2) I(|d| < 1)
\end{equation}
We separately fit $\hat\mu_s(x,y)$ and $\hat\mu_t(t)$ and include a weighting term $m_0$ to enable the model
to downweight this background intensity:
\begin{equation}
\hat\mu(x,y,t) = m_0 \cdot \hat\mu_s(x,y) \hat\mu_t(t)
\end{equation}
We performed a sensitivity analysis in Section \label{section:smoothing-kernel} to investigate the impact
of the kernel bandwidth.

In order to learn the parameters of the spatiotemporal Hawkes process, we use
Bayesian inference implemented with Markov Chain Monte Carlo methods to find
the posterior distribution of the parameters, where we place weakly informative 
priors \cite{gelman2008weakly} on them:
\begin{itemize}
\item $m_0 \sim \mathcal{N}_+(0,1)$
\item $\theta \sim \mathcal{N}_+(0,10)$
\item $\sigma \sim \mathcal{N}_+(0,10)$ where distance is measured in kilometers
\item $\omega \sim \mathcal{N}_+(0,10)$ where time is measured in 1/days
\end{itemize}
with $\mathcal{N}_+$ denoting the Normal distribution truncated to be positive.

We coded our model in the probabilistic programming language Stan \cite{stan-software:2016},
which implements Hamiltonian Monte Carlo sampling to efficiently explore the posterior distribution
ove all of the parameters of our model. We ran 4 chains for 2,000 iterations each. Convergence
diagnostics are discussed in Section \ref{section:mcmc}.

\subsection{Dealing with holidays}
\label{section:holidays}
\begin{figure}[ht!]
	\centering
            \includegraphics[width=.45\linewidth]{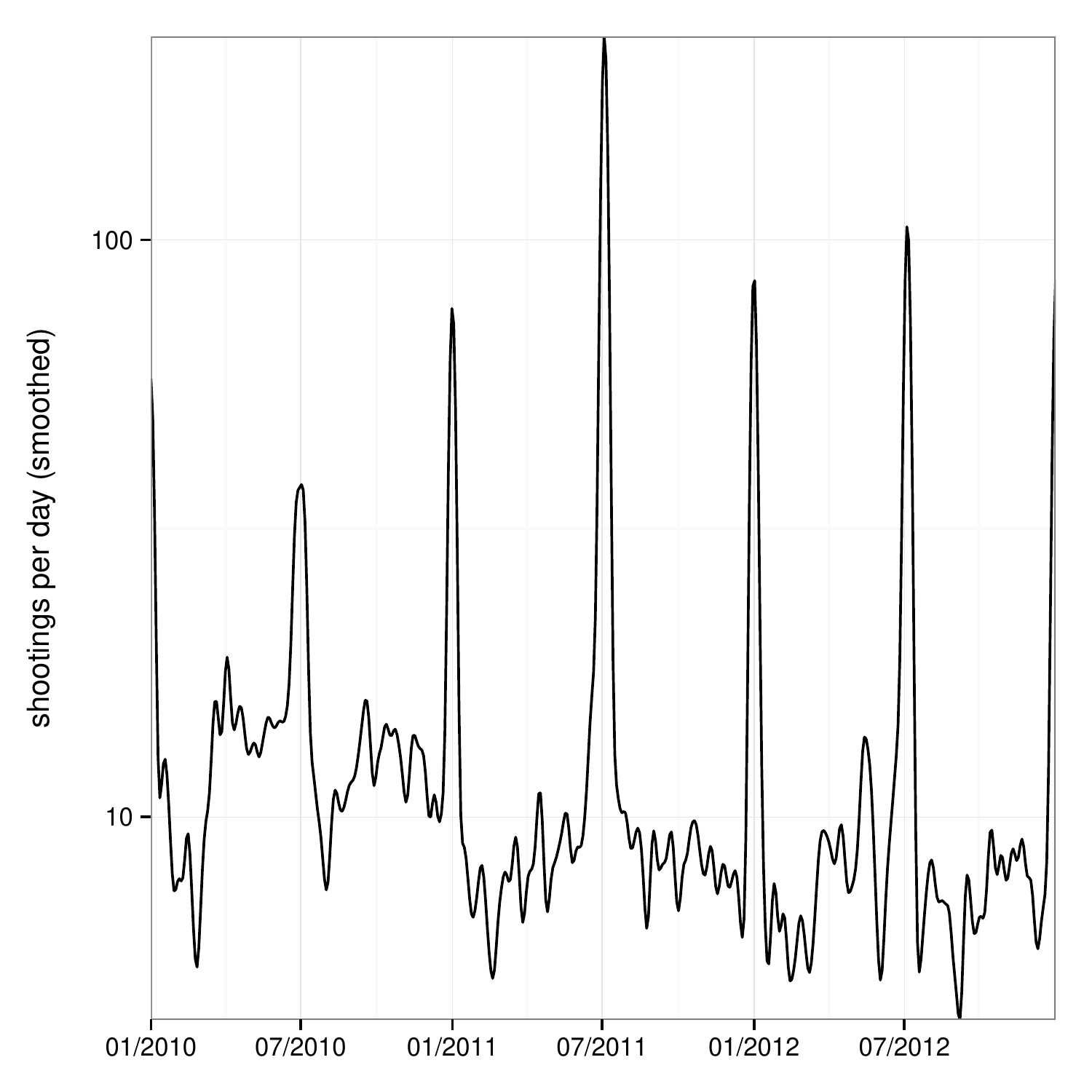}%
            \includegraphics[width=.45\linewidth]{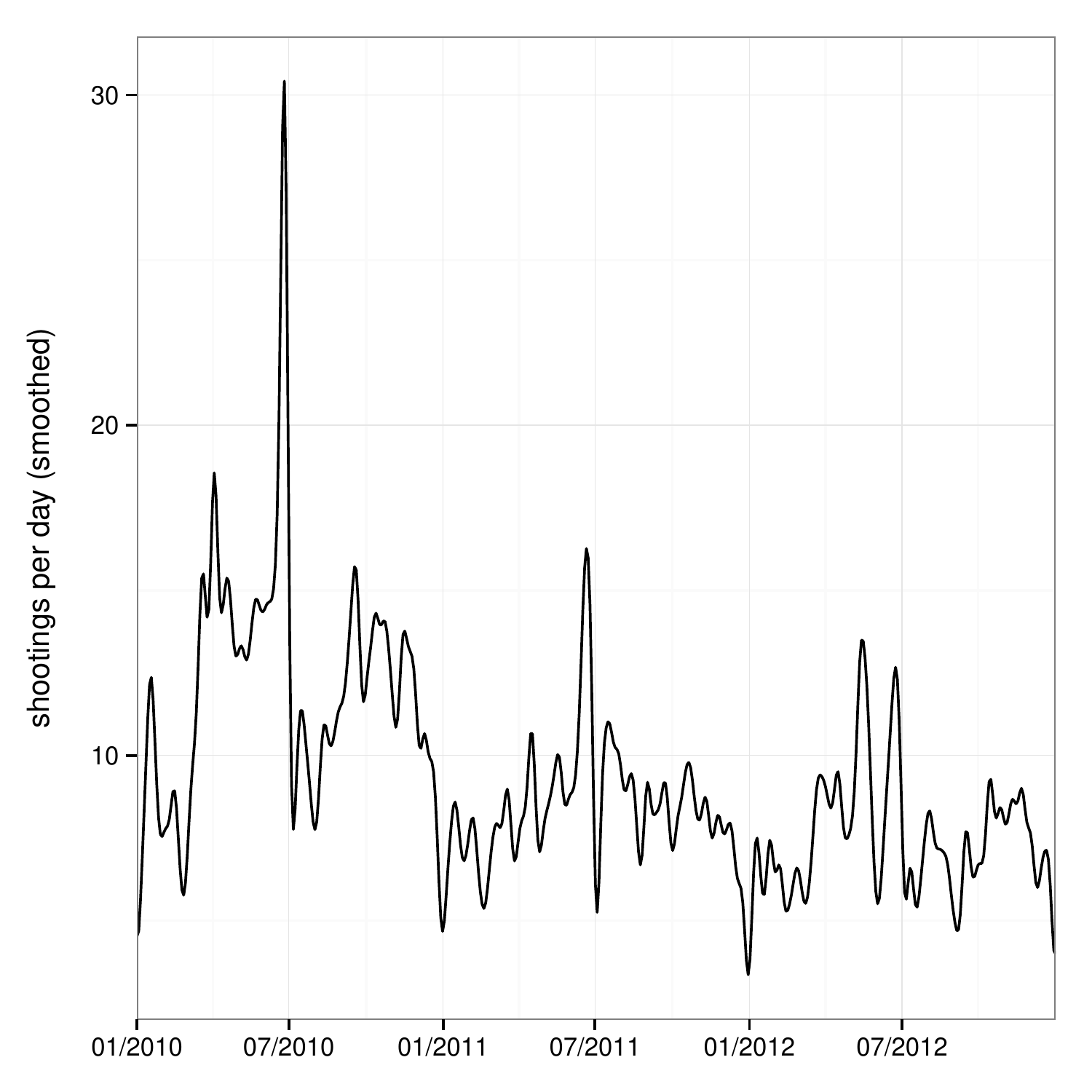}
        \caption{Smoothed time series of shots fired in Washington D.C., 2011, comparing the original data (left) to the data after removing July 1st-6th and December 29th-January 2nd for each year (right).}\label{fig:temporal-intensity}
\end{figure}

As shown in Figure \ref{fig:temporal-intensity} (left), many extra shootings are observed in our dataset on the days
before and after New Year's Eve and the Fourth of July.  There were 5,653 total shootings on these days, accounting for 36\% of all shootings.
We considered these shootings to be ``celebratory'' and thus
removed July 1st-6th and December 29th-January 2nd from our dataset for each year. The time series without these extra shootings
is shown in Figure \ref{fig:temporal-intensity} (right). 
We performed a sensitivity analysis to see whether our main findings were robust to including the holidays. The posterior over the lengthscales was slightly longer---19 minutes and 210 meters and the parameter $\theta$ (which characterizes the number of additional shootings we expect to see after a single shooting) was higher: 0.43 (versus 0.13 in our main model) and correspondingly $m_0$ the weight placed on the underlying intensity was 0.57 (versus 0.87 in our main model). All of these findings make sense, as the holiday shootings are by their nature spatiotemporally clustered and the model represents this by increasing the weight and spatiotemporal extent of the self-excitatory component.

\subsection{Crime Offense Reports Data} 
 Conventional reported crime data on firearm assaults, as seen in Figure 5, manifests a similar spatial distribution and temporal distribution as AGLS-detected gunshots.  However, the intensity levels are markedly different. As a result of these intensity differences between measures, heterogeneity in the intensity of incident clusters is much more visible in the AGLS spatial distribution than the conventional firearm assault distribution. In addition, as seen in the bottom right graph in Figure 5, firearm assault data lags AGLS data by several hours. A lag of several minutes would be consistent with existing knowledge on reporting practices, but a lag of hours likely reflects a recording practice associated with duty shifts or a similar process. For this reason, 911 calls for service data on ``shots fired'' will provide a better source for estimating spatiotemporal models if AGLS data is not present and binning is observed in publicly reported crimes data.
\par
To verify that our spatiotemporal models could be used on 911 calls for service data, we refit our final model using the 911 calls for service data for ``sounds of gunshots'' for an overlapping temporal period beginning in 2011 and ending in 2013. The spatial lengthscale was 221 [214,228] meters. The temporal lengthscale was 9 [8.4,9.4] minutes. And the self-exciting paramters, corresponding to the fraction of events attributable to the conditional as opposed to background intensity, was 0.15 [0.14,0.16]. These results closely matched those obtained using AGLS data. The degree of similarity is likely a product of the high fraction of gunshot events reported to police both by civilians and the AGLS system. In cities with more limited reporting or deployment, models fit on different data sources may not yield as consistent results.
\begin{figure}[h]
	\centering
		\centering
		\includegraphics[width=1.05\linewidth]{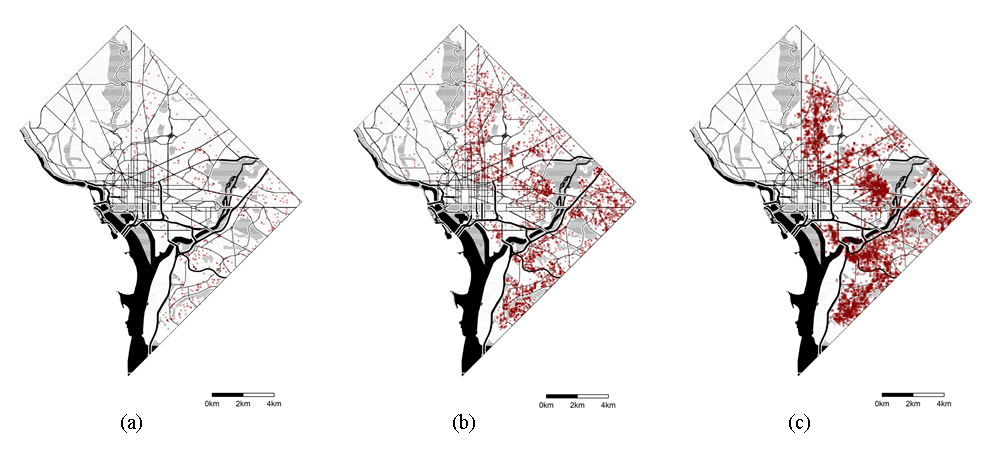}%
	
		\setcounter{subfigure}{0} \renewcommand{\thesubfigure}{d}
    \begin{subfigure}[t]{.45\linewidth}
		\includegraphics[width=1\linewidth]{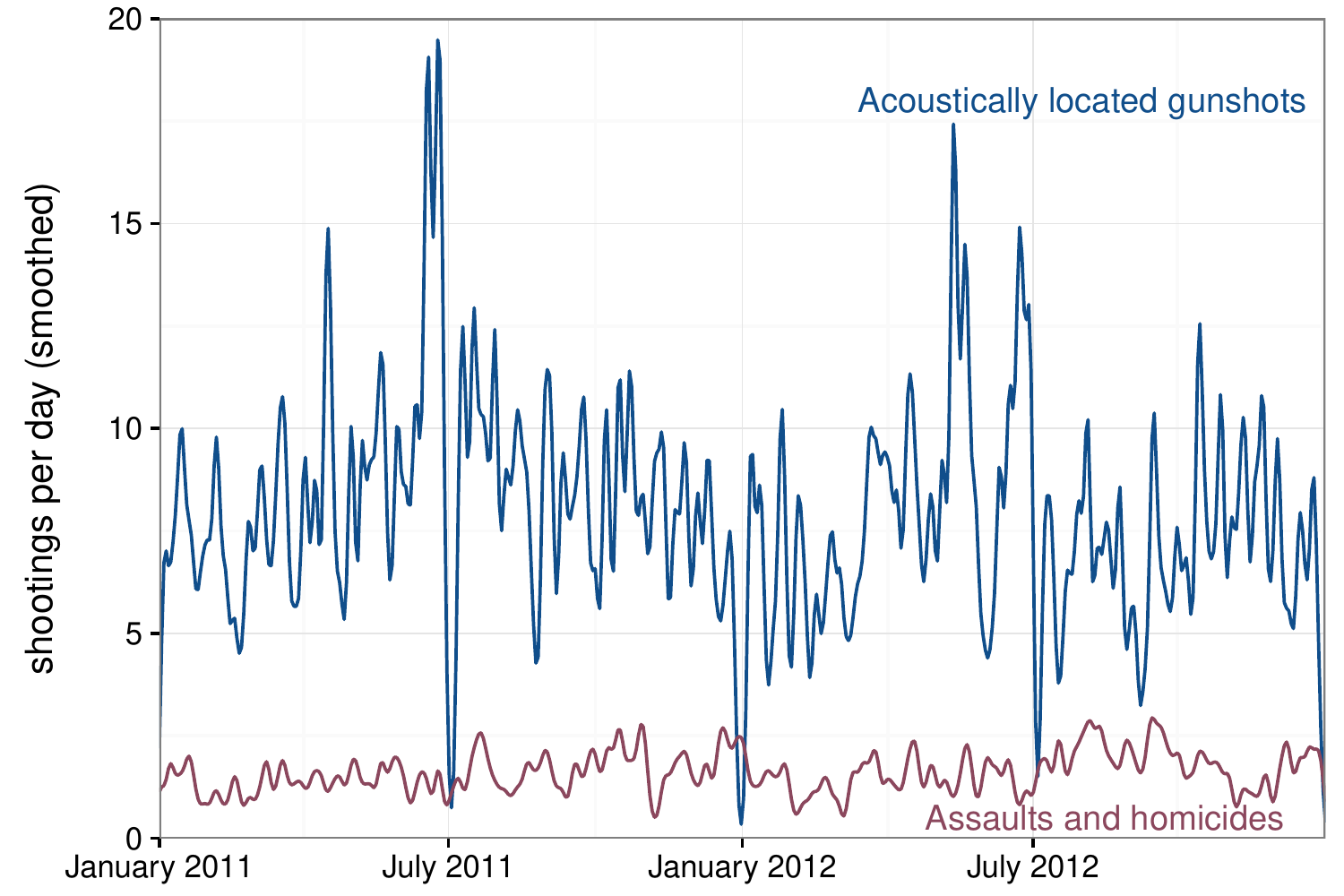}
		\caption{}\label{fig:shots-fired}
	\end{subfigure}
		\setcounter{subfigure}{0} \renewcommand{\thesubfigure}{e}
	\begin{subfigure}[t]{.45\linewidth}
		\includegraphics[width=1\linewidth]{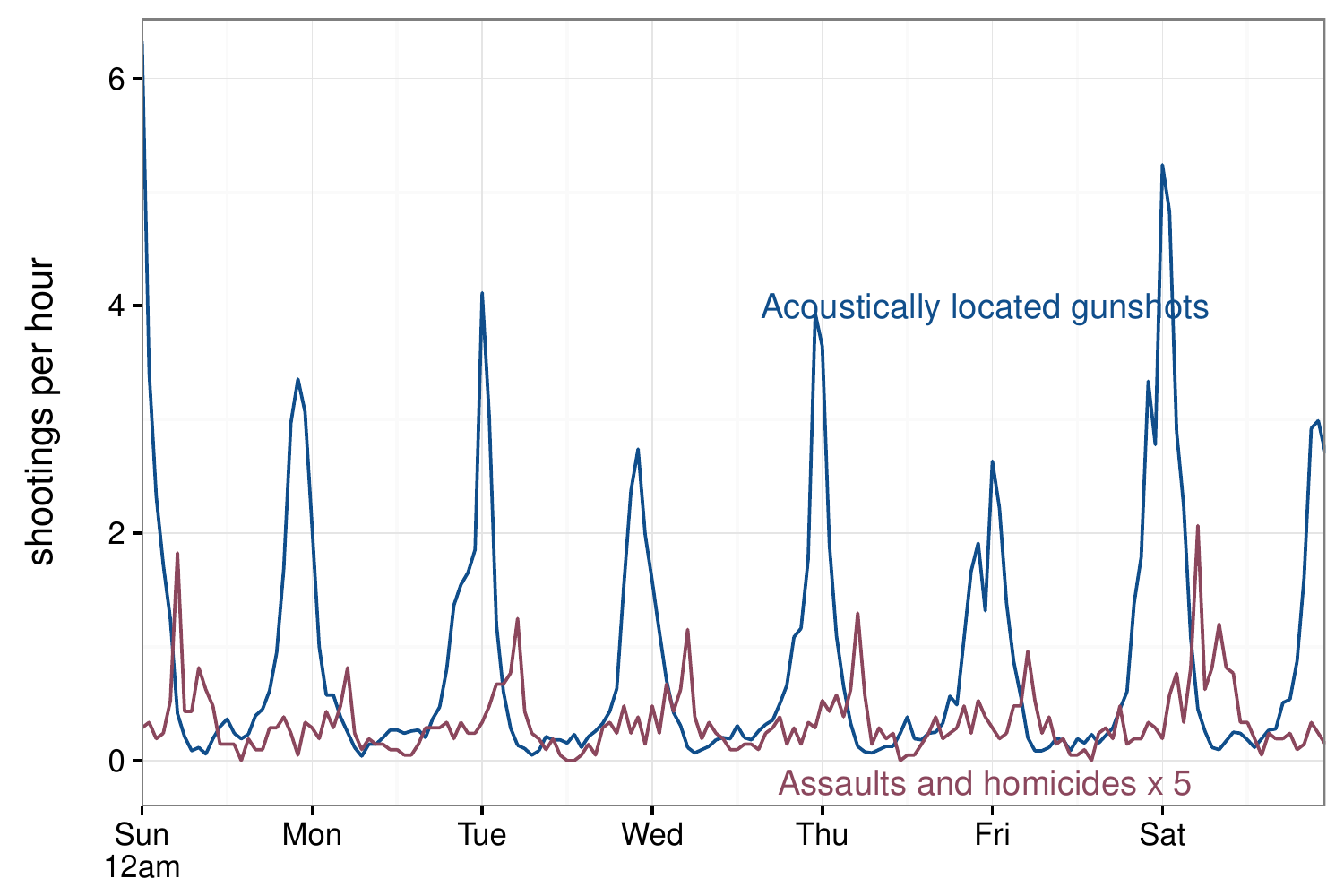}
		\caption{}\label{fig:shots-fired}
	\end{subfigure}
	\caption{Spatial and Temporal Distributions of Firearm Assaults and Shots Fired in Washington D.C., 2011. The spatial distribution of aggravated assaults (a), 911 "shots fired" calls (b), and acoustically detected gunshots (c) is shown subfigures (a-c). Monthly and day-of-week temporal distributions are shown in subfigures (d-e).}\label{fig:intensity}
\end{figure}

\subsection{Sensitivity analysis for smoothing kernel bandwidth}
\label{section:smoothing-kernel}
We investigate the choice of the bandwidth for the smoothing kernel for estimating the underlying intensity $\hat \mu(x,y,t)$.
For spatial bandwidths in $[0.5, 1, 1.6]$ km and temporal bandwidths in $[0.5, 1, 7, 14]$ days we fit our model using only 
data from 2011-12. The learned spatial lengthscales were between 130 and 210 meters, matching our main model's posterior (126 meters), while
the learned temporal lengthscales were between 12 minutes and 20 minutes, slightly longer than in our main model (10 minutes).

\subsection{Sensitivity analysis for inclusion of duplicate events}
\label{section:duplicates}
In our main analysis, we merged shootings 
which occurred within 1 minute and 100 meters together, treating them as a single shooting (after also removing holidays). We thus removed 415 shootings which we considered to be duplicates (4\% of our sample).
As a sensitivity analysis, we fit our model to the original dataset, without holidays, but with each of the near-repeats included. All of our findings were consistent with our main findings: the lengthscale was 100 meters in space (slightly shorter than the main model) and 11 minutes in time, and $m_0$, the weight placed on the endogeneous component was 0.86, while $\theta$ was 0.14. This
underscores the very local and limited extent to which we are able to find evidence of diffusion in our dataset.

\subsection{Traceplots and convergence diagnostics}
\label{section:mcmc}
\begin{figure}[h]
    \centering
                \includegraphics[width=.95\linewidth]{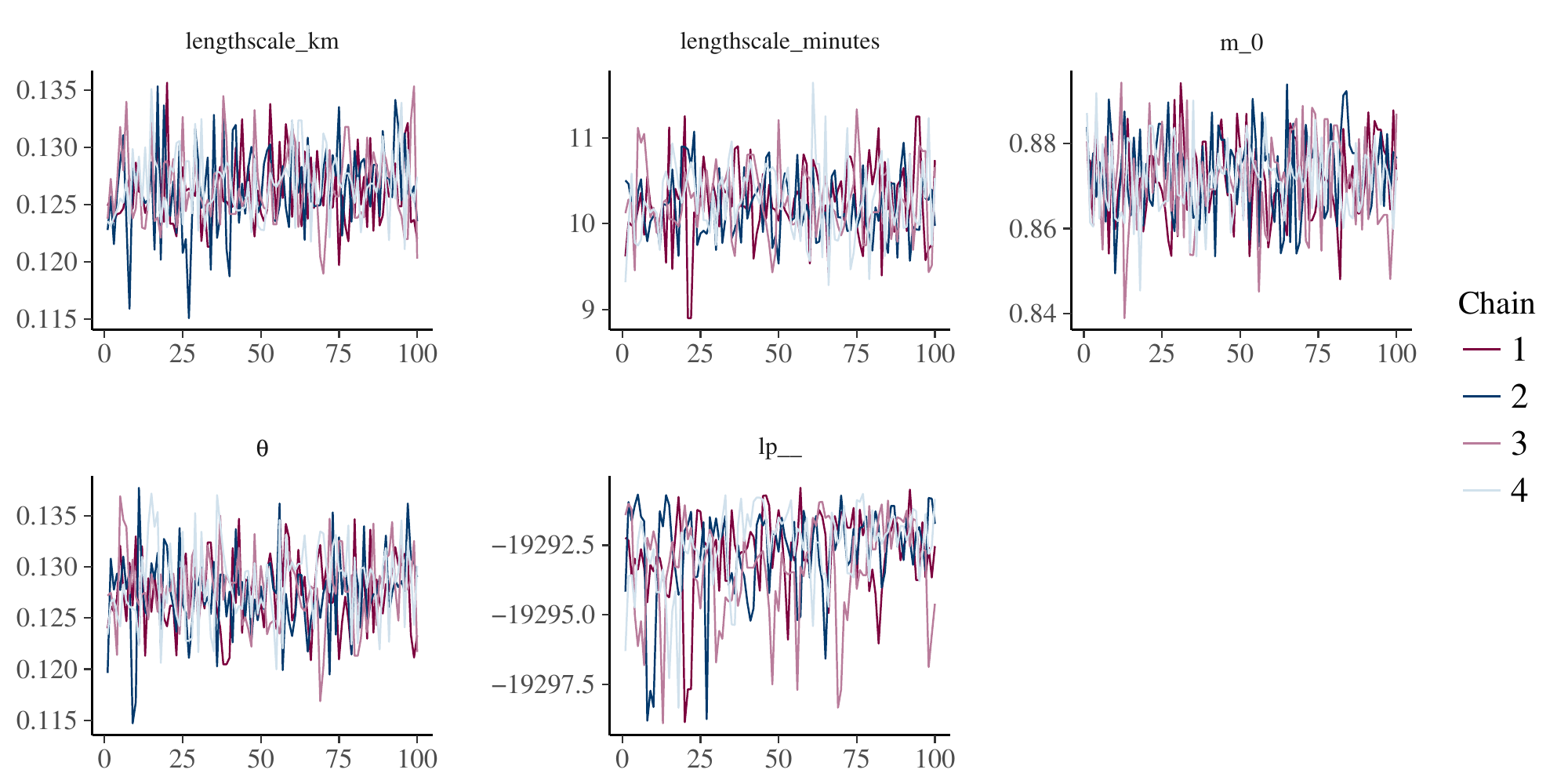}
                \caption{Traceplots for the parameters of the model fit with HMC, showing good convergence and mixing.}
                \label{fig:traceplots}
\end{figure}

We implemented the model in Section \ref{section:hawkes} in the probabilistic programming
language Stan \cite{stan-software:2016}, which uses Hamiltonian Monte Carlo to approximately
sample from the posterior of the distribution over the parameters. We ran 4 chains for 200
iterations each, discarding the first 100 draws as burn-in.
We show traceplots in Figure \ref{fig:traceplots}. The Gelman-Rubin statistic $\hat R$ is very close to 1 for every parameter. Thus we conclude that the chains have mixed and converged, and we are sampling from the true posterior of the model.

\end{document}